\documentclass{iaus}
\usepackage{graphicx}

\def\arcsec{\nobreak{$''$}}

\title[The Impact of Nuclear Star Formation on Gas Inflow to AGN] %% give here short title %%
{The Impact of Nuclear Star Formation on Gas Inflow to AGN}

\author[R. Davies et al.]   %% give here short author list %%
{R.I.~Davies$^1$,
E.~Hicks$^1$,
M.~Schartmann$^{1,2}$,
R.~Genzel$^1$,
L.J.~Tacconi$^1$,
H.~Engel$^1$,
A.~Burkert$^{1,2}$,
M. Krause$^{1,2}$,
A.~Sternberg$^3$,
F.~Mueller~S\'anchez$^4$,
\and 
W.~Maciejewski$^5$
}

\affiliation{$^1$Max-Planck-Institut f\"ur extraterrestrische Physik,
  Garching, Germany\\
$^2$Universit\"ats-Sternwarte M\"unchen, Germany\\
$^3$School of Physics and Astronomy, Tel Aviv University, Tel Aviv, Israel\\
$^4$Instituto de Astrofisica de Canarias, La Laguna, Tenerife, Spain\\
$^5$Astrophysics Research Institute, Liverpool John Moores University, UK}

\pubyear{2009}
\volume{267}  %% insert here IAU Symposium No.
\pagerange{119--126}
\jname{Co-Evolution of Central Black Holes and Galaxies}
\editors{B.M.\ Peterson, R.S.\ Somerville, \& T.\ Storchi-Bergmann, eds.}
\begin{document}

\maketitle

\begin{abstract}
Our adaptive optics observations of nearby AGN at spatial resolutions as
small as 0.085\arcsec\ show strong evidence for recent, but no longer
active, nuclear star formation. 
We begin by describing observations that highlight two contrasting methods by
which gas can flow into the central tens of parsecs.
Gas accumulation in this region will inevitably lead to a starburst, and
we discuss the evidence for such events.
We then turn to the
impact of stellar evolution on the further inflow of gas by combining
a phenomenological approach with analytical modelling and 
hydrodynamic simulations. These complementary perspectives paint a picture
in which all the processes are ultimately regulated by the mass accretion
rate into the central hundred parsecs, and the ensuing starburst that
occurs there. The resulting supernovae delay accretion by generating a
starburst wind, which leaves behind a clumpy interstellar medium. This
provides an ideal environment for slower stellar outflows to accrete
inwards and form a dense turbulent disk on scales of a few parsecs. 
Such a scenario may resolve the discrepancy between the larger scale 
structure seen with adaptive optics and the small scale structure seen 
with VLTI.
\end{abstract}

\firstsection % if your document starts with a section,
\section{Introduction}

Many studies of active galactic nuclei (AGN), particularly those
concerned with understanding the co-evolution of black holes and their
host galaxies through cosmic time, rely on observations of quasars.
The reason is simply that these are the most luminous of such objects and
can be studied at all redshifts.
While they form an important component of the AGN population, the
luminosity function (\cite[Hao et al. 2005]{hao05}) indicates that the
number density of lower luminosity AGN is orders of magnitude greater.
This is particularly true in the local universe because of the
effects of downsizing, which means that at low redshift black hole
growth occurs primarily in AGN with black hole masses
$<10^8$\,M$_\odot$ (\cite[Best et al. 2005]{bes05}) and Seyfert-like
luminosities (\cite[Hasinger et al. 2005]{has05}).
As a result, with only a few exceptions such as Mkn\,231
(\cite[Davies et al. 2004]{dav04b}), detailed studies of nearby AGN
have to focus on the lower luminosity AGN such as Seyferts.

This has important implications. Although most quasar activity may be
fuelled via mergers of gas rich galaxies, Seyfert nuclei typically
reside in (preferentially early type) spiral galaxies 
(\cite[Ho 2008]{ho08}).
As a result, Seyfert activity is likely to be fuelled via secular
processes associated with disk evolution such as those outlined by
\cite{shl90} and \cite{wad04}.
One obvious part of the overall scheme is the role played by bars,
which are known to drive gas inwards from the large-scale disk.
Yet despite the many attempts to find a correlation between the
presence of a bar and an AGN, the statistical outcome is, at best, for
only a marginal link between these phenomena.
This, perhaps, should not be surprising.
Bars drive large amounts of gas inwards (of order
0.1\,M$_\odot$\,yr$^{-1}$, \cite[Regan \& Teuben 2004]{reg04}), 
working on spatial scales of kpcs and timescales of Gyrs.
In contrast, a Seyfert nucleus requires typically
$\lesssim0.01$\,M$_\odot$\,yr$^{-1}$ of gas 
(\cite[Jogee et al. 2006]{jog06}), and the relevant temporal and
spatial scales are Myr and pc.
The two regimes are orders of magnitude different in every respect.
Thus, while it is well understood how bars can create a gas reservoir
and a starburst in the central kiloparsec (and there is plenty of observational
evidence to support the link between bars and both of these
phenomena:
\cite[Sakomoto et al. 1999]{sak99},
\cite[Laurikainen et al. 2004]{lau04},
and \cite[Jogee et al. 2005]{jog05} are a few exmaples among many),
this is only a single possible step in the process that leads to AGN
fuelling.

It is now understood that the properties of the central kpc -- the
circumnuclear region -- can play an important role in driving gas further
inwards.
But although observations have yielded many tantalising results, reaching
definitive conclusions is difficult.
For example, \cite{mar03} looked at the detailed circumnuclear
dust morphologies of matched samples of active and inactive galaxies.
One of their results was that AGN are only found in galaxies
that exhibit dusty spiral patterns.
But they also found that spiral patterns of any sort were associated
equally often with active and inactive galaxies.
This, again, is likely a result of the differing spatial and temporal
scales of the phenomena.
The aspect of time variability is one of the issues addressed by NUGA
(\cite[Garc\'ia-Burillo et al. 2007]{gar07}), a study of the
distribution and kinematics of the molecular gas in nearby low
luminosity AGN.
One of their key results was that gravitational torques can not only
drive gas in to form a circumnuclear ring, but also tend to drive
gas out from the nuclear region towards the ring.
This led \cite{gar05} to propose a scenario in which further inward
flow was possible episodically if two conditions were fulfilled: the
bar should be weakened by dynamical feedback, and the viscosity should
increase due to a sufficiently steep density gradient in the
circumnuclear ring.

These examples show that there is still much we do not understand
about the conditions under which gas can be transported inward from kpc
scales to the nuclear region, and it 
is one of the topics addressed here: we describe two highly
contrasting case studies where gas is being driven inwards to the
central tens of parsecs.
We also show evidence that, once there, the gas is likely to trigger a
starburst.
And we discuss the impact of the starburst on the fate of
the gas, and a possible link to the sub-parsec scales.

\section{Watching Gas Flow Inwards}

In this section, we describe contrasting cases where gas can be
observed as it flows in towards an AGN. 
Although we present only two examples here, analyses of non-circular
motions and gas inflow mechanisms on similar 
scales have been performed for other galaxies as well:
NGC\,7469 (\cite[Davies et al. 2004]{dav04a}),
NGC\,6951 (\cite[Storchi-Bergmann et al. 2007]{sto07}), 
NGC\,4051 (\cite[Riffel et al. 2008]{rif08}),
NGC\,7582 (\cite[Riffel et al. 2009]{rif09}),
NGC\,4151 (\cite[Storchi-Bergmann et al. 2009]{sto09}).

In one case, 
NGC\,1068 (\cite[Mueller S\'anchez et al. 2009]{mue09}), 
the inflow rate is prodigious, due to chance events, and probably
cannot be sustained. In the other, 
NGC\,1097 (\cite[Davies et al. 2009]{dav09}), 
the inflow is much more
moderate and appears to be regulated in a way that is sustainable for
timescales of a Gyr.
Intriguingly, in neither case do we see gas flowing all the way down
to the AGN: the inflow is terminated at spatial scales of a few to
tens of parsecs, where it is instead expected to fuel a starburst.

\subsection{Gas inflow in NGC\,1068}

Among the nearby AGN we have studied 
(\cite[Davies et al. 2007, Hicks et al. 2009]{dav07,hic09}), NGC\,1068
is rather unusual. 
The H$_2$ emission in the central 250\,pc originates in an expanding
off-centre ring or shell with particularly bright and massive clumps
around the north east side.
Filaments of gas extend from the ring at a radius of about 30\,pc to
the AGN on both sides.
\cite{mue09} has modelled the morphology and kinematics of the
filaments, finding that the only way to simultaneously account for
both constraints is if the filaments trace gas that is falling almost
directly in towards the AGN.
These models indicate that the infall timescale is about 1.3\,Myr.
One of the filaments lies across the front of the AGN, suggesting that
inelastic collisions may allow the gas to settle on scales of a few
parsecs.

Estimating the mass of gas in the filaments is extremely difficult.
\cite{mue09} used a variety of methods to derive a mass of
$\sim2\times10^7$\,M$_\odot$, although with about a factor of 3
uncertainty.
Thus the inflow rate to the central few parsecs is of order
15\,M$_\odot$\,yr$^{-1}$.
This is a remarkably high rate, 2--3 orders of magnitude greater than
that needed to power the AGN itself.
And because the characteristics of the gas in the central region are
so unusual, and the inflow appears to be due to chance combination of
circumstances, such a rapid inflow is probably unsustainable.

\subsection{Gas inflow in NGC\,1097}

\begin{figure}[t]
\begin{center}
\includegraphics[width=13cm]{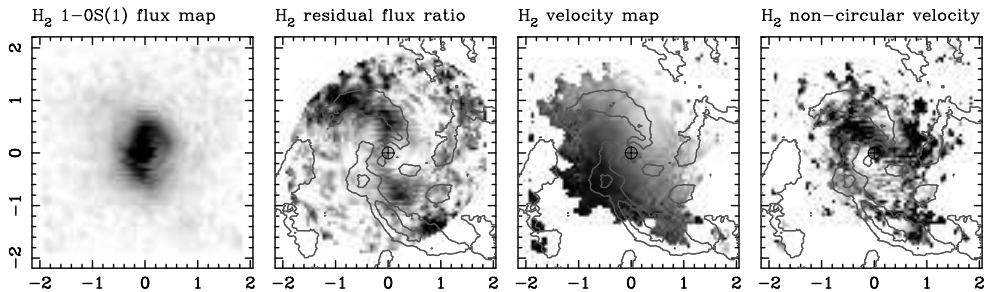}
\end{center}
\caption{Flux and velocity maps of the 1-0\,S(1) H$_2$ emission line
  in NGC\,1097. 
Far and centre left: flux map and residual after subtracting
  elliptical isophotes, showing the 3 photometric arms.
Centre and far right: velocity field and residual after subtracting
  circular motions, showing the 2 kinematic arms.
Adapted from \cite{dav09}.}
\label{davies:fig:ngc1097}
\end{figure}

NGC\,1097 exhibits many classical features of disk galaxies: a large
scale bar, a circumnuclear ring and starburst, an inner spiral, and a
low luminosity AGN. 
Its one peculiarity is that the inner spiral has 3 photometric arms
(\cite[Prieto et al. 2005]{pri05}).
Integral field observations (\cite[Davies et al. 2009]{dav09})
showed that this pattern is due to obscuration of stellar light by gas
and dust in the molecular arms.
Intriguingly, as seen in Fig.~\ref{davies:fig:ngc1097}, the gas
kinematics reveal a strong non-circular 
velocity residual in the form of a 2-arm spiral -- as expected
from linear theory which indicates that the projected line-of-sight
velocity pattern of an $m$-arm spiral is an ($m-1$)-arm spiral.
The properties of the arms allow one to derive an inflow rate along
them of $\sim1.2$\,M$_\odot$\,yr$^{-1}$.
However, hydrodynamical simulations of nuclear spirals show that there
is also significant outflow between the arms in the low density
regions (gas from the arms overshoots the nucleus
and continues as a diverging outflow).
In the simulations, the net inflow rate is a factor 20 less;
applying this to NGC\,1097 means it could be as little as
0.06\,M$_\odot$\,yr$^{-1}$.

Given that there is of order $10^8$\,M$_\odot$ of gas within a radius
of 700\,pc, such a modest inflow rate would require 1.8\,Gyr to drain
the gas reservoir in the circumnuclear region.
If this gas is replenished by the ring, which itself is fed by the
large-scale bar, then the inflow could in principle continue even
longer.
Thus the inner spiral in NGC\,1097 appears to be driving gas inwards
at a sustainable rate.

As for NGC\,1068, the gas is not reaching the AGN. Instead the spiral
arms appear to terminate in the central 10--20\,pc, at which point
the gas dispersion increases dramatically 
(\cite[Davies et al. 2007, Hicks et al. 2009]{dav07,hic09}).
If the gas piles up here, it will inevitably lead to a starburst --
and a nuclear starburst has been reported 
(\cite[Storchi-Bergmann et al. 2005, Davies et al. 2007]{sto05,dav07}).
The estimated gas inflow rate could lead to episodic starbursts similar to
this one every 20--150\,Myr.

\section{Nuclear Starbursts}

We have seen in the previous section how, at least for these 2 cases,
gas flows into the central few to tens of parsecs.
If it is stalled here, one would naturally expect it to
trigger a starburst.
Evidence for starbursts on these scales was presented by \cite{dav07}, who
used near infrared integral field spectroscopy combined with adaptive
optics to achieve high spatial resolution.
Using the spectral information, these authors were able to separate
out the non-stellar continuum (hot dust emission) associated with the
AGN, and show that in every case the stellar continuum itself was
spatially resolved with a typical size scale $<$50\,pc.
Furthermore, they showed that this was a distinct stellar population,
younger and dynamically cooler than the bulge: photometrically, from an excess
in the stellar continuum above an $r^{1/4}$ law fitted to 1--2\arcsec\
scales and extrapolated inwards; and kinematically, from a drop in the stellar
dispersion on the same spatial scales.
Constraints on stellar population synthesis models provided by
several independent diagnostics indicated that the starburst ages were young,
lying in the range 10--300\,Myr.
One important piece of evidence was the remarkably low equivalent width of
the Br$\gamma$ line, which shows unambiguously that the star formation
has ceased: i.e. although still young, these are post-starbursts.
Intriguingly, starburst models imply that during the short (perhaps
only 10\,Myr) actively star forming phase, the luminosity of the
nuclear starburst would have been an order of magnitude greater than
it is now -- with a star formation rate per unit area comparable to that
in ULIRGs.
This high intensity appears to make sense in terms of the high gas
mass surface density which, from the Kennicutt-Schmidt law, is
expected to lead to high star formation rates.

\begin{figure}[t]
\begin{center}
\includegraphics[width=11cm]{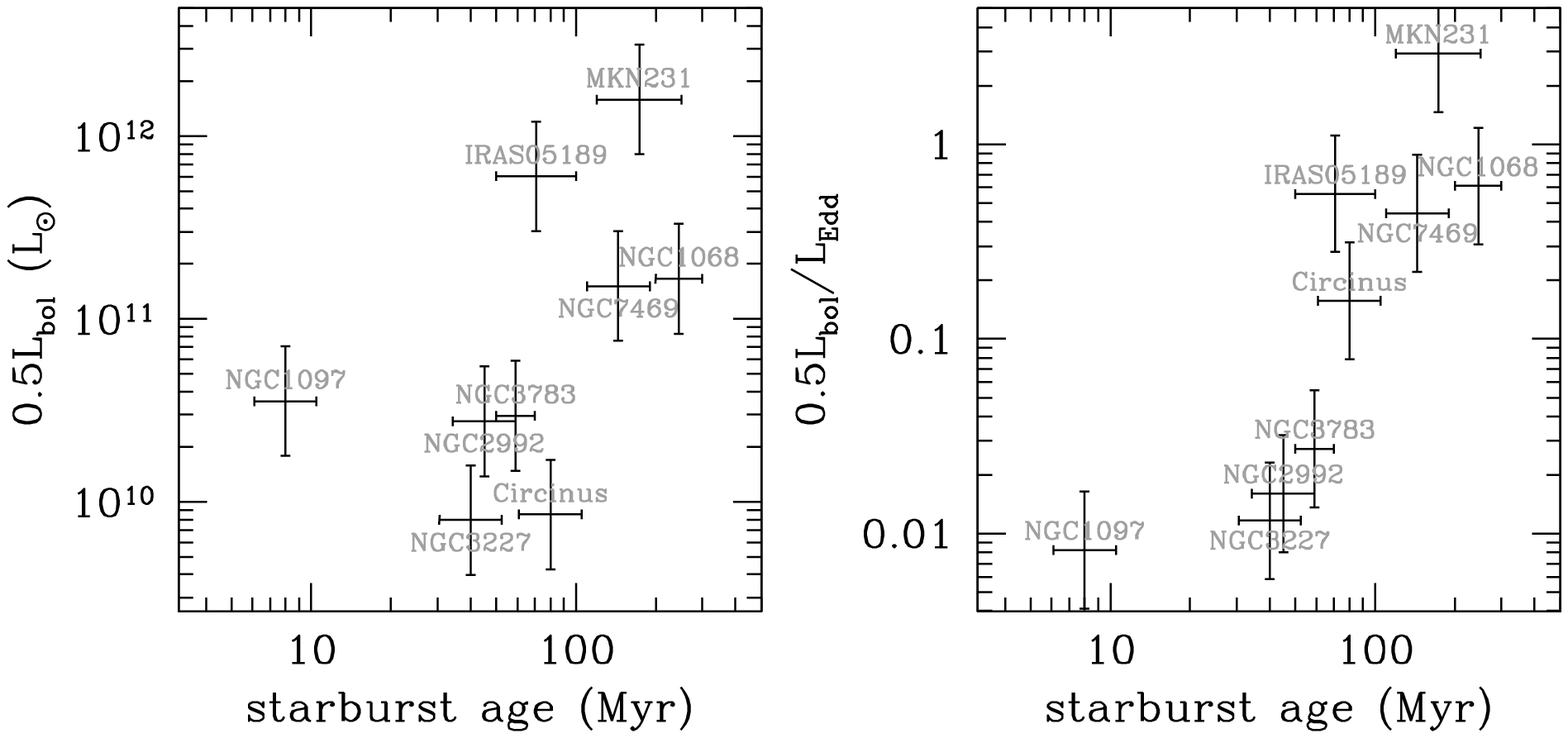}
\end{center}
\caption{Relation between the characteristic age of most recent starburst
  and the AGN luminosity and accrection rate. This figure is adapted
  from \cite{dav07} and includes an additional point for NGC\,2992
  from \cite{fri09}.}
\label{davies:fig:sb_age}
\end{figure}

More pertinent to the issue of gas inflow, the authors showed that
there appears to be a relation between the characteristic age of the
most recent episode of star formation and the luminosity of, or accretion
rate onto, the AGN (Fig.~\ref{davies:fig:sb_age}).
This figure suggests that there is a delay of 50--100\,Myr between the
short duration starburst and the onset of accretion onto the AGN.
This timescale is interesting because it corresponds to the age at
which, for a short burst, one expects the type\,{\sc ii} supernovae
phase to be finishing.
This topic is discussed in more detail in
Section~\ref{davies:sec:outflows}, which addresses the impact of
stellar winds and supernovae on gas inflow and outflow, and how these
might provide a link between phenomena observed with adaptive optics on
$\sim$10\,pc scales and those seen with infrared interferometric techniques on
0.1--1\,pc scales.

\section{Stellar Outflows and the Torus: Linking 10pc and 1pc Scales}
\label{davies:sec:outflows}

The data presented in \cite{dav07} and \cite{hic09} show that
nuclear star formation is occuring on the same spatial scales, up to
a few tens of parsecs, as the gas distribution; and that on these
scales the two components have similar kinematics.
This means the stars and gas must be mixed.
It is inevitable that the star formation will have some impact on the gas,
and this was argued from a phenomenological approach by
\cite{dav07}.
Hydrodynamical simulations of the impact of stellar evolution, such as
those presented in \cite{sch09} give a clearer picture.
To understand this in the context of Seyfert nuclei, these simulations
have been scaled to match a typical starburst there, as described by
Schartmann et al. (these proceedings): 
M$_{\rm BH}\sim10^7$\,M$_\odot$, 
M$_{\rm stars}\sim2\times10^8$\,M$_\odot$ out to 50\,pc,
a dispersion of $\sigma*\sim100$\,km\,s$^{-1}$ as well as rotation,
and a mass loss rate $\sim0.1$\,M$_\odot$\,yr$^{-1}$ corresponding
to an age of 50\,Myr but which of course would vary with time.
The simulations show that there are two, fairly well defined, regimes
separated by a stellar outflow speed of about 250\,km\,s$^{-1}$.
Fast outflows (which could represent winds from OB stars, as well as
type\,{\sc ii} and type\,{\sc i} supernovae) lead to a starburst wind
which efficiently removes the diffuse ISM.
Slower outflows (which represent, for example, planetary nebulae or winds from
AGB stars) lead to a large number of clumps that can accrete
efficiently to smaller scales, giving rise to a dense and turbulent
disk with a radius of 0.5--1\,pc.
Taken together, these imply that there is a phase in a starburst when
accretion to smaller scales is possible and efficient. 
For a short starburst, this phase would begin after about 50\,Myr, once
there are no more OB stars or type\,{\sc ii} supernovae;
and it would end after about 300\,Myr once type\,{\sc i} supernovae
start to appear.

The size scale of the central disk is similar to that of the
compact 8--12\,$\mu$m structure 
in NGC\,1086 (the galaxy to which the simulation is
scaled) by \cite{rab09} using MIDI on the VLTI; and within a factor of
a few of the compact structure seen in Circinus by \cite{tri07} using
the same instrument.
As already suggested by \cite{dav08}, these simulations imply that
stellar outflows are a
link between the gas structure seen on scales of 10--50\,pc using
adaptive optics, and those seen on 0.1--1\,pc scales using
infrared interferometry.
This scenario has a number of similarities to the simulations of
\cite{wad09}.
Although the initial conditions and the distribution of energy input
are rather different, the result is a vertically thick (up to 10\,pc
from the disk plane) distribution at large radii, and in the inner few
parsec a compact turbulent disk.

\begin{figure}[t]
\begin{center}
\includegraphics[width=12cm]{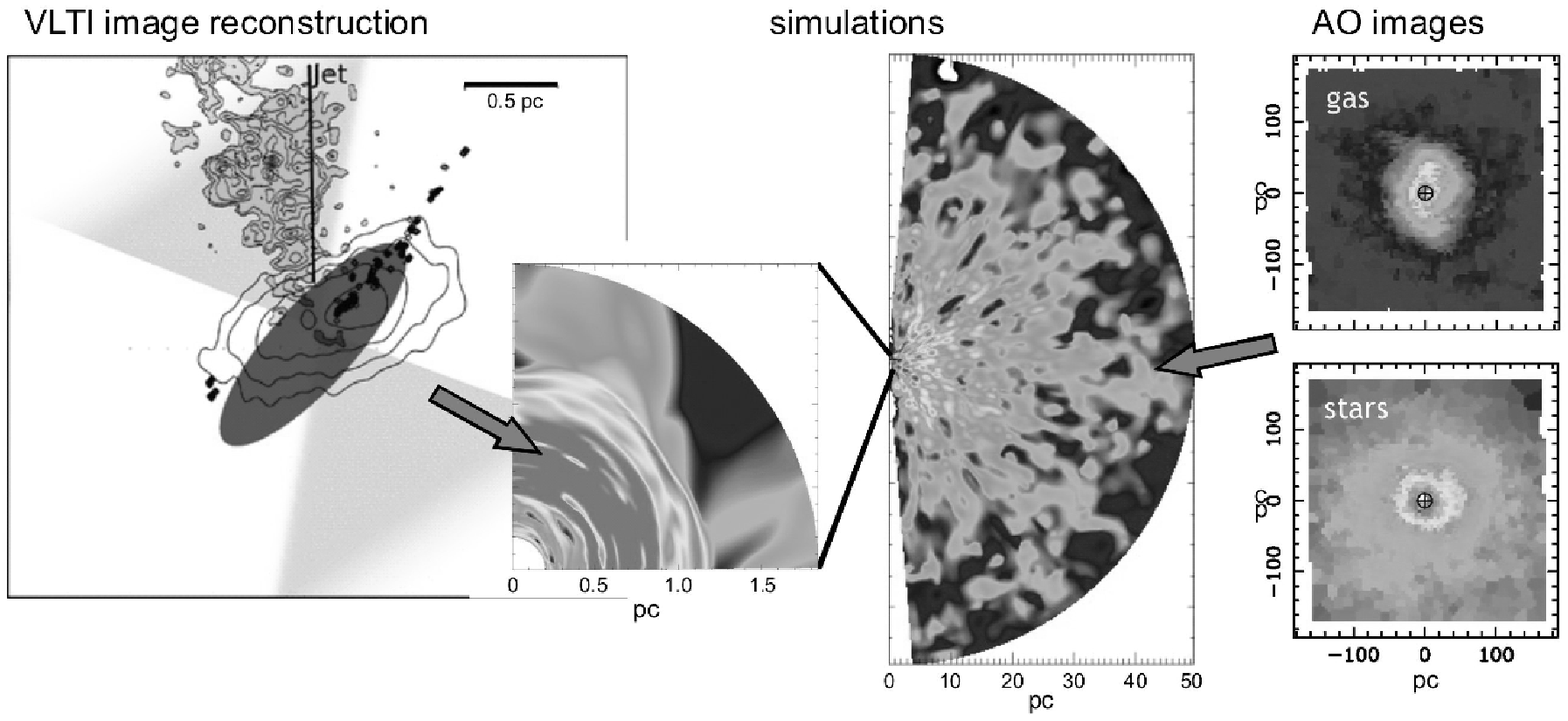}
\end{center}
\caption{Composite figure showing how hydrodynamical simulations of
  stellar outflows might provide a link between the 10\,pc
  and 1\,pc scale structure within the torus. Panels are taken from
  \cite{rab09}, \cite{hic09}, \cite{dav09}, and simulations similar to those in
  \cite{sch09} but scaled to a typical Seyfert nucleus. This figure is
  illustrative, and the individual panels do not
  necessarily refer to the same specific object.}
\label{davies:fig:links}
\end{figure}

If one considers either of these structures to be associated with the
molecular obscuring torus, then one must also conclude that they both
are. As illustrated in Fig.~\ref{davies:fig:links} and stated by
\cite{hic09}, this implies that the torus
comprises multiple components each of which fulfils different roles;
that it must be a star forming torus; 
and, because the starbursts are episodic, that it must be a dynamical entity.
What we see on 10--50\,pc scales is the global structure, a
diffuse and clumpy envelope where there are short repeated bursts of star
formation. The slow stellar winds accrete down to smaller $\sim$1\,pc
scales where 
they form the more compact structure, a dense and turbulent disk.
The evolution of this disk -- in terms of further star formation and
gas inflow -- is addressed by Schartmann et al. (these proceedings).

One final point worth noting is the similarity between these
simulations and those performed for the Galactic Center, on scales
about 100 times smaller, by \cite{cua06} and \cite{cua08}.
Both simulations were performed within a 0.5\,pc volume in which the
stars, their orbits, and their outflows, matched as precisely as
possible the known configuration at that time.
In the earlier version, where there were both slow $\sim$200\,km\,s$^{-1}$
and fast $\sim$700\,km\,s$^{-1}$ stellar winds, the result was that
much of the ISM was blown out  but that clumps of cooler dense gas
accreted to form a compact disk that extended out to $\sim$0.04\,pc.
In the later version, the updated stellar wind speeds were all fast,
and nearly the entire ISM was blown out.
It is reassuring that qualitatively, these simulations yield the same
result as those of \cite{sch09} even though the scales are drastically
different,
because in both cases it is the same processes (i.e. stellar outflows)
that govern the evolution of the region.

%\section{Maintaining the High Dispersion}

\section{Conclusions}

We have presented a brief review of the current status of our work on
the impact that star formation has on gas inflow to Seyfert nuclei.
We are not yet able to follow gas all the way down to the AGN itself,
and instead we have focussed on spatial scales from a few hundred
parsecs down to the central parsec. Our main conclusions are:

\begin{itemize}
\item
We have seen 2 contrasting methods of gas inflow that feed a nuclear
starburst. While showing how gas might reach the central few to tens
of parsecs, this also raises questions about the sustainability of
inflows, how far in they reach, and whether it is possible to define
what a typical inflow mechanism or rate is.

\item
There are short intense starbursts in the central tens of parsecs around
Seyfert nuclei. In general, starbursts will be an inevitable
consequence of gas inflow.

\item
Stellar outflows play a key role in the evolution of the ISM on these
scales. Specifically, fast outflows tend to blow out the diffuse ISM,
while slow outflows can lead to efficient accretion of gas onto
smaller scales. This may be a link between what is seen on
10--50\,pc scales with adaptive optics, and what is observed on
0.1--1\,pc scales with infrared interferometry.

\item
The molecular obscuring torus most likely consists of multiple
structures on different scales that fulfil different
roles. 
Since the state of the torus is strongly affected by the episodic star
formation that occurs within it, we need to think of it as a
dynamic entity.

\end{itemize}


\begin{thebibliography}{}

\bibitem[Best et al. (2005)]{bes05}
Best P., Kauffmann G., Heckman T., Birnchmann J., Charlot S.,
Iverzi\'c \v Z., White S., 2005, 
MNRAS, 362, 25

\bibitem[Cuadra et al. (2006)]{cua06}
Cuadra J., Nayakshin S., Springel V., Di Matteo T., 2006,
MNRAS, 366, 358

\bibitem[Cuadra et al. (2008)]{cua08}
Cuadra J., Nayakshin S., Martins F., 2008,
MNRAS, 383, 458

\bibitem[Davies et al. (2004a)]{dav04a} % NGC7469
Davies R., Tacconi L., Genzel R., 2004a,
ApJ, 602, 148

\bibitem[Davies et al. (2004b)]{dav04b} % Mkn231
Davies R., Tacconi L., Genzel R., 2004b,
ApJ, 613, 781

\bibitem[Davies et al. (2007)]{dav07}
Davies R., Mueller Sanchez D., Genzel R., Tacconi L., Hicks E.,
Friedrich S., Sternberg A., 2007,
ApJ, 671, 1388

\bibitem[Davies et al. (2008)]{dav08}
Davies R., 2008,
NewAR, 52, 307

\bibitem[Davies et al. (2009)]{dav09}
Davies R., Maciejewski W., Hicks E., Tacconi L., Genzel R., Engel H.,
2009, 
ApJ, 702, 114

\bibitem[Friedrich et al. (2009)]{fri09}
Friedrich S., Davies R., Hicks E., Engel H., Mueller S\'anchez F.,
Genzel R., Tacconi L., 2009,
A\&A, submitted

\bibitem[Garc\'ia-Burillo et al. (2005)]{gar05}
Garc\'ia-Burillo S., Combes F., Schinnerer E., Boone F., Hunt K., 2005
A\&A, 441, 1011

\bibitem[Garc\'ia-Burillo et al. (2007)]{gar07}
Garc\'ia-Burillo S., Combes F., Usero A., Grac\'ia-Carpio J., 2007,
NewAR, 51, 160

\bibitem[Hasinger et al.(2005)]{has05}
Hasinger G., Miyaji T., Schmidt M., 2005,
A\&A, 441, 417

\bibitem[Hao et al. (2005)]{hao05}
Hao L., et al., 2005,
AJ, 129, 1795

\bibitem[Hicks et al. (2009)]{hic09}
Hicks E., Davies R., Malkan M., Genzel R., Tacconi L.,
Mueller S\'anchez F., Sternberg A., 2009,
ApJ, 696, 448

\bibitem[Ho (2008)]{ho08}
Ho L., 2008,
ARA\&A, 46, 475

\bibitem[Jogee et al. (2005)]{jog05}
Jogee S., Scoville N., Kenney J., 2005,
ApJ, 630, 837

\bibitem[Jogee (2006)]{jog06}
Jogee S., 2006,
{\em Lecture Notes in Physics}, 693, 143

\bibitem[Laurikainen et al. (2004)]{lau04}
Laurikainen E., Salo H., Buta R., 2004,
ApJ, 607, 103

\bibitem[Martini et al. (2003)]{mar03}
Martini P., Regan M., Mulchaey J., Pogge R., 2003,
ApJ, 589, 774

\bibitem[Mueller S\'anchez et al. (2009)]{mue09}
Mueller S\'anchez F., Davies R., Genzel R., Tacconi L., Eisenhauer F.,
Hicks E., Friedrich S., Sternberg A., 2009,
ApJ, 691, 749

\bibitem[Prieto et al.(2005)]{pri05}
Prieto A., Maciejewski W., Reunanen J., 2005,
AJ, 130, 1472

\bibitem[Raban et al. (2009)]{rab09}
Raban D., Jaffe W., R\"ottgering H., Meisenheimer K., Tristram K.,
2009,
MNRAS, 394, 1325

\bibitem[Regan \& Teuben (2004)]{reg04}
Regan M., Teuben P., 2004,
ApJ, 600, 595

\bibitem[Riffel et al. (2008)]{rif08}
Riffel R., Storchi-Bergmann T., Winge C., McGregor P., Beck T.,
Schmitt H., 2008,
MNRAS, 385, 1129

\bibitem[Riffel et al. (2009)]{rif09}
Riffel R., Storchi-Bergmann T., Dors O., Winge C., 2009,
MNRAS, 393, 783

\bibitem[Sakamoto et al. (1999)]{sak99}
Sakamoto K., Okumura S., Ishizuki S., Scoville N., 1999,
ApJ, 525, 691

\bibitem[Schartmann et al. (2009)]{sch09}
Schartmann M., Meisenheimer K., Klahr H., Camenzind M., Wolf S.,
Henning T., 2009,
MNRAS, 393, 759

\bibitem[Shlosman et al. (1990)]{shl90}
Shlosman I., Begelman M., Frank J., 1990,
Nature, 345, 679

\bibitem[Storchi-Bergmann et al. (2005)]{sto05}
Storchi-Bergmann T., Nemmen R., Spinelli P., Eracleous M., Wilson A.,
Filippenko A., Livio M., 2005,
ApJ, 624, L13

\bibitem[Storchi-Bergmann et al. (2007)]{sto07}
Storchi-Bergmann T., Dors O., Riffel R., Fathi K., Axon D., 
Robinson A., Marconi A., \"Ostlin G., 2007,
ApJ, 670, 959

\bibitem[Storchi-Bergmann et al. (2009)]{sto09}
Storchi-Bergmann T., Sim\~oes-Lopes R., McGregor P., Riffel R., 
Beck T., Martini P., 2009,
MNRAS, accepted

\bibitem[Tristram et al. (2007)]{tri07}
Tristram K., et al., 2007,
A\&A, 474, 837

\bibitem[Wada (2004)]{wad04}
Wada K., 2004, 
in {\em Carnegie Observatories Astrophysics Series, Vol. 1:
  Coevolution of Black Holes and Galaxies}, ed. Ho L. (CUP)
p. 186

\bibitem[Wada et al. (2009)]{wad09}
Wada K., Papadopoulos P., Spaans M., 2009,
ApJ, 702, 63

\end{thebibliography}
\end{document}